\documentclass{desyprocA4}

\begin{document}
\title{Single top quark polarization at $O(\alpha_{s})$ in $t \bar t$
  production\hbox to0pt{\vtop to 0pt{\kern-96pt\normalsize\rm\hbox to0pt{\hss
  MZ-TH/12-41}\hbox to0pt{\hss August 2012}\normalsize\kern96pt\vss\ }\hss}\\
at a polarized linear $e^+e^-$ collider}

\author{{\slshape S.~Groote$^1$, J.G.~K\"orner$^2$,
  B.~Meli\'c$^3$, S.~Prelovsek$^4$}\\[1ex]
$^1$Loodus- ja Tehnoloogiateaduskond, F\"u\"usika Instituut,
  Tartu \"Ulikool, Riia~142, EE--51014 Tartu, Estonia \\
$^2$Institut f\"ur Physik der Johannes-Gutenberg-Universit\"at,
Staudinger Weg 7, D--55099 Mainz, Germany \\
$^3$ Rudjer Bo\v skovi\'c Institute, Theoretical Physics Division,
Bijeni\v cka c. 54, HR--10000 Zagreb, Croatia\\
$^4$ Physics Department at University of Ljubljana
and Jozef Stefan Institute, SI--1000 Ljubljana, Slovenia  }
\contribID{xy}

\confID{1964}  
\desyproc{DESY-PROC-2010-01}
\acronym{PLHC2010} 
\doi  

\maketitle

\begin{abstract}
We present a detailed investigation of the NLO polarization of the top quark
in $t \bar t$ production at a polarized linear $e^+e^-$ collider with 
longitudinally polarized beams. By appropiately
tuning the polarization of the beams one can achieve close to maximal values
for the top quark polarization over most of the forward hemisphere for a large
range of energies. This is quite welcome since the rate is largest in the
forward hemisphere. One can also tune the beam polarization to obtain close to
zero polarization over most of the forward hemisphere.
\end{abstract}

\section{\label{sec1}Introductory remarks}
The top quark is so heavy that it keeps its polarization at production when it 
decays since $\tau_{\rm hadronization}\gg\tau_{\rm decay}$. One can test the
Standard Model (SM) and/or non-SM couplings through polarization measurements
involving top quark decays (mostly  $t\to b+W^+$). New observables involving
top quark polarization can be defined such as 
$\langle\vec{P}_t\cdot\vec{p}\rangle$ (see e.g.\ Refs.~\cite{Christova:1997by,%
Fischer:1998gsa,Fischer:2001gp,Groote:2006kq,AguilarSaavedra:2010nx,
Drobnak:2010ej}).
It is clear that the analyzing power of such observables is largest for large
values of the polarization of the top quark. This calls for large top quark
polarization values. One also wants a control sample with small or zero top
quark polarization. Near maximal and minimal values of top quark polarization
at a linear $e^+e^-$ collider can be achieved in $t \bar t$ production by 
appropiately tuning the
longitudinal polarization of the beam polarization~\cite{Groote:2010zf}. At
the same time one wants to keep the top quark pair production cross section
large. It is a fortunate circumstance that all these goals can be realized at
the same time. A polarized linear $e^+e^-$ collider may thus be viewed as
a rich source of close to zero and close to $100\%$ polarized top quarks.

\begin{figure}[ht]
\begin{center}
\includegraphics[width=0.6\textwidth]{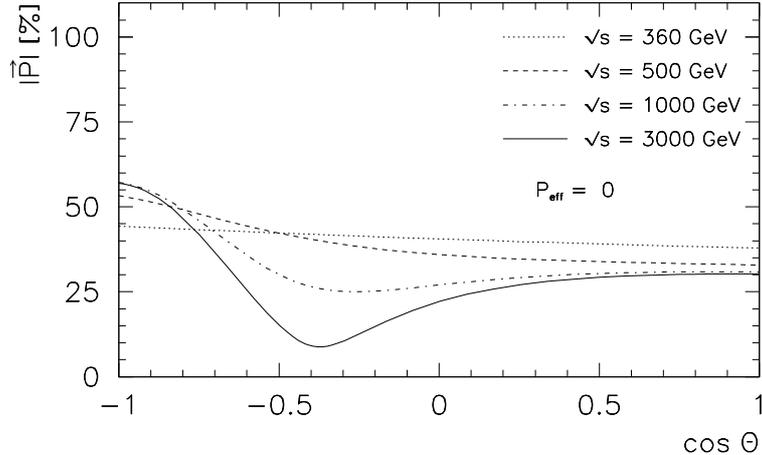} 
\end{center}
\vspace*{-0.5cm}
\caption{\label{fig:zeropol}Magnitude of NLO top quark polarization for zero 
beam polarization}
\end{figure}

Let us remind the reader that the top quark is polarized even for zero beam 
polarization through vector--axial vector interference effects
$\sim v_ea_e,\,v_ea_f,\,v_fa_e,\,v_fa_f$, where 
\begin{eqnarray}\label{ewcouplings}
v_e,a_e\quad&:& \mbox{electron current coupling}\nonumber\\
v_f,a_f\quad&:& \mbox{top quark current coupling}
\end{eqnarray}
In Fig.~\ref{fig:zeropol} we present a NLO plot of the $\cos\theta$ dependence 
of the zero beam polarization top quark polarization for different
characteristic energies at  $\sqrt{s}=360$\,GeV (close to threshold),
$\sqrt{s}=500$\,GeV (ILC phase 1), $\sqrt{s}=1000$\,GeV (ILC phase 2) and
$\sqrt{s}=3000$\,GeV (CLIC). 

\section{\label{sec2}Top quark polarization at threshold
  and in the high energy limit}
The polarization of the top quark depends on the c.m.\ energy $\sqrt{s}$, the
scattering angle $\cos\theta$, the electroweak coupling coefficients $g_{ij}$
and the effective beam polarization $P_{\rm eff}$, i.e.\ one has
\begin{equation}\label{vecpol}
\vec{P}=\vec{P}\,(\sqrt{s},\cos\theta, g_{ij},P_{\rm eff})\,,
\end{equation}
where the effective beam polarization appearing in Eq.~(\ref{vecpol}) is given
by~\cite{MoortgatPick:2005cw}
\begin{equation}\label{peff}
P_{\rm eff}=\frac{h_--h_+}{1-h_-h_+}\,.
\end{equation}
and where $h_-$ and $h_+$ are the longitudinal polarization of the electron
and positron beams $(-1<h_{\pm}<+1)$, respectively. Instead of the nonchiral
electroweak couplings $g_{ij}$ one can alternatively use the chiral
electroweak couplings $f_{mm'}$ ($m,m'=L,R$) introduced in
Refs.~\cite{Parke:1996pr,Kodaira:1998gt}. The relations between the two sets of
electroweak coupling  coefficients can be found in Ref.~\cite{Groote:2010zf}.
In this report we shall make use of both sets of coupling parameters.

For general energies the functional dependence in Eq.~(\ref{vecpol}) is not 
simple. Even if the electroweak couplings $g_{ij}$ are fixed, one remains with 
a three-dimensional parameter space $(\sqrt{s},\cos\theta,P_{\rm eff})$. Our
strategy is to discuss various limiting cases for the Born term polarization
and then to investigate how the limiting values extrapolate away from these
limits. In particular, we exploit the fact that, in the Born term case,
angular momentum conservation (or $m$-quantum number conservation) implies
$100\%$ top quark polarization at the forward and backward points for the
$(e^-_Le^+_R)$ and $(e^-_Re^+_L)$ beam configurations.

In this section we discuss the behaviour of $\vec{P}$ at nominal threshold 
$\sqrt{s}=2m_t$ ($v=0$) and in the high energy limit $\sqrt{s}\to\infty$
($v\to 1$). At threshold and at the Born term level one has
\begin{equation}\label{thresh}
\vec{P}_{\rm thresh}=\frac{P_{\rm eff}- A_{LR}}{1-P_{\rm eff}A_{LR}}
\,\,\,\hat{n}_{e^-}\,,
\end{equation}
where $A_{LR}$ is the left--right beam polarization asymmetry
$(\sigma_{LR}-\sigma_{RL})/(\sigma_{LR}+\sigma_{RL})$ and $\hat{n}_{e^-}$ is a
unit vector pointing into the direction of the electron momentum. We use a
notation where $\sigma(LR/RL)=\sigma(h_-=\mp 1;h_+=\pm 1)$. In terms of the
electroweak coupling parameters $g_{ij}$, the nominal polarization asymmetry
at threshold $\sqrt{s}=2m_t$ is given by
$A_{LR}=-(g_{41}+g_{42})/(g_{11}+g_{12})=0.409$. Eq.~(\ref{thresh}) shows
that, at threshold and at the Born term level, the polarization $\vec{P}$ is
parallel to the beam axis irrespective of the scattering angle and has maximal
values $|\vec{P}|=1$ for both $P_{\rm eff}=\pm1$ as dictated by angular
momentum conservation. Zero polarization is achieved for
$P_{\rm eff}=A_{LR}=0.409$.

In the high energy limit the polarization of the top quark is purely
longitudinal, i.e.\ the polarization points into the direction of the top
quark. At the Born term level one finds 
$\vec{P}(\cos\theta)=P^{(\ell)}(\cos\theta)\cdot\hat{p_t}$ with
\begin{equation}\label{helimit}
P^{(\ell)}(\cos\theta)
  =\frac{(g_{14}+g_{41}+P_{\rm eff}(g_{11}+g_{44}))(1+\cos\theta)^2
  +(g_{14}-g_{41}-P_{\rm eff}(g_{11}-g_{44})(1-\cos\theta)^2}
  {(g_{11}+g_{44}+P_{\rm eff}(g_{14}+g_{41}))(1+\cos\theta)^2
  +(g_{11}-g_{44}-P_{\rm eff}(g_{14}-g_{41}))(1-\cos\theta)^2}\,.
\end{equation}
In the same limit, the electroweak coupling coefficients appearing in 
Eq.~(\ref{helimit}) take the numerical values $g_{11}=0.601$, $g_{14}=-0.131$,
$g_{41}=-0.201$ and $g_{44}=0.483$. For $\cos\theta=\pm1$ and
$P_{\rm eff}=\pm1$ the top quark is $100\%$ polarized as again dictated by 
angular momentum conservation. The lesson from the threshold and high energy 
limits is that large values of the polarization of the top quark close to 
$|\vec{P}|=1$ are engendered for large values of the effective beam
polarization parameter close to $P_{\rm eff}=\pm1$.

Take, for example, the forward--backward asymmetry which is zero at threshold,
and large and positive in the high energy limit. In fact, from the numerator
of the high energy formula Eq.~(\ref{helimit}) one calculates
\begin{equation}
A_{FB}=\frac34\,\,\frac{g_{44}+P_{\rm eff}g_{14}}{g_{11}+P_{\rm eff}g_{41}}
\,=\,0.61\,\frac{1-0.27P_{\rm eff}}{1-0.33P_{\rm eff}}\,.
\end{equation}
The forward-backward asymmetry is large and only mildly dependent on 
$P_{\rm eff}$. More detailed calculations show that the strong forward
dominance of the rate sets in rather fast above
threshold~\cite{Groote:2010zf}. This is quite welcome since the forward
region is also favoured from the polarization point of view.

As another example take the vanishing of the polarization which, at threshold,
occurs at $P_{\rm eff}=0.409$. In the high energy limit, and in the forward
region where the numerator part of Eq.~(\ref{helimit}) proportional to
$(1+\cos\theta)^2$ dominates, one finds a polarization zero at
$P_{\rm eff}=(g_{14}+g_{41})/(g_{11}+g_{44})=0.306$. The two values of
$P_{\rm eff}$ do not differ much from another.

\section{\label{sec3}Overall rate and left-right (LR)
  and right-left (RL) rates}
\begin{figure}[t]
\begin{center}
\includegraphics[width=0.6\textwidth]{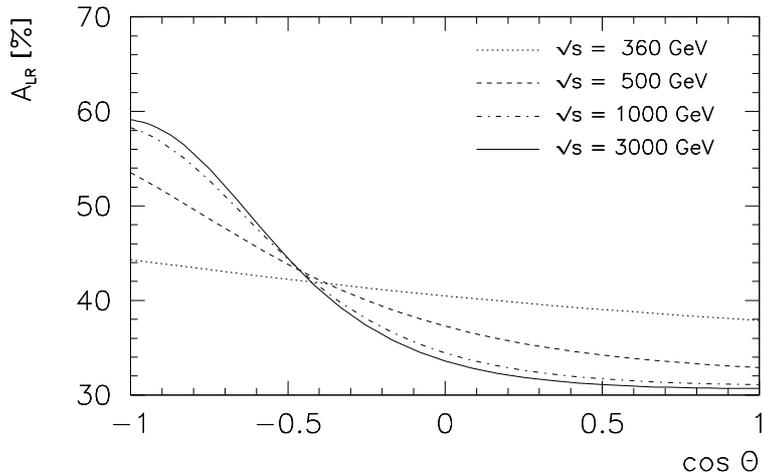}
\end{center}
\vspace*{-0.5cm}
\caption{\label{fig:asymrl}NLO left--right polarization asymmetry $A_{LR}$ for
  $\sqrt s=360$, $500$, $1000$, and $3000$\,GeV}
\end{figure}

The overall rate $\sigma$ for partially longitudinal polarized beam production
can be composed from the LR rate $\sigma_{LR}$ and the RL rate $\sigma_{RL}$
valid for $100\%$ longitudinally polarized beams. The notation is such that LR
and RL refer to the $(e^-_Le^+_R)$ and $(e^-_Re^+_L)$ longitudinal
polarization configurations, respectively. The relation
reads~\cite{Artru:2008cp}
\begin{eqnarray}\label{rate}
\frac{d\sigma}{d\cos\theta}
  &=&\frac{1-h_-}2\frac{1+h_+}2\frac{d\sigma_{LR}}{d\cos\theta}
  +\frac{1+h_-}2\frac{1-h_+}2\frac{d\sigma_{RL}}{d\cos\theta}\nonumber\\
  &=&\frac14(1-h_-h_+)\Big(\frac{d\sigma_{LR}+d\sigma_{RL}}{d\cos\theta}
  -P_{\rm eff}\frac{d\sigma_{LR}-d\sigma_{RL}}{d\cos\theta}\Big)\,.
\end{eqnarray}
Using the left--right polarization asymmetry
\begin{equation}
A_{LR}=\frac{d\sigma_{LR}-d\sigma_{RL}}{d\sigma_{LR}+d\sigma_{RL}}
\end{equation}
one can rewrite the rate~(\ref{rate}) in the form
\begin{equation}\label{ratealr}
\frac{d\sigma}{d\cos\theta}=\frac14(1-h_-h_+)
  \frac{d\sigma_{LR}+d\sigma_{RL}}{d\cos\theta}\Big(1-P_{\rm eff}A_{LR}\Big)\,.
\end{equation}
The differential rate $d\sigma/d\cos\theta$ carries an overall helicity
alignment factor $(1-h_-h_+)$ which enhances the rate for negative values of
$h_-h_+$. Also, Fig.~\ref{fig:asymrl} shows that $A_{LR}$ varies in the range
between $0.30$ and $0.60$ which leads to a further rate enhancement from the
last factor in Eq.~(\ref{ratealr}) for negative values of $P_{\rm eff}$. 

Let us define reduced LR and RL rate functions $D_{LR/RL}$ by writing
\begin{equation}
\frac{d\sigma_{LR/RL}}{d\cos\theta}
  =\frac{\pi\alpha^2v}{3s^2}D_{LR/RL}(\cos\theta)
\end{equation}
such that, in analogy to Eq.~(\ref{rate}),
\begin{equation}\label{ovrate}
D=\frac14(1-h_-h_+)\left(D_{LR}+D_{RL}-P_{\rm eff}(D_{LR}-D_{RL})\right)\,.
\end{equation} 
In the next step we express the reduced rate functions through a set of
independent hadronic helicity structure functions. For the LR reduced rate
function one has
\begin{eqnarray}
2D_{LR}(\cos\theta)
  &=&\frac38(1+\cos^2\theta)
  \left((f_{LL}^2+f_{LR}^2)H_U^1+2f_{LL}f_{LR}H_U^2\right)\nonumber\\&&
  +\frac34\sin^2\theta
  \left((f_{LL}^2+f_{LR}^2)H_L^1+2f_{LL}f_{LR}H_L^2\right)\nonumber\\&&
  +\frac34\cos\theta(f_{LL}^2-f_{LR}^2)H_F^4
\end{eqnarray}
and accordingly for $D_{RL}$ with $f_{LL}\to f_{RR}$ and $f_{LR}\to f_{RL}$. 

At NLO one has $H_a^j=H_a^j({\it Born\/})+H_a^j(\alpha_s)$. The radiatively
corrected structure functions $H_a^j(\alpha_s)$ are listed in
Ref.~\cite{Groote:2010zf}. If needed they can be obtained from S.G.\ or
B.M.\ in Mathematica format. For the non-vanishing unpolarized Born term
contributions  $H_a^j({\it Born\/})$ one obtains (see e.g.\
Ref.~\cite{Groote:2008ux,Groote:2010zf})
\begin{eqnarray}\label{HUL1}
H_U^1({\it Born\/})=2N_cs(1+v^2),&&
H_L^1({\it Born\/})=\ H_L^2({\it Born\/})=N_cs(1-v^2),\nonumber\\[3pt]
H_U^2({\it Born\/})=2N_cs(1-v^2),&&
H_F^4({\it Born\/})=4N_csv.
\end{eqnarray}

Following Refs.~\cite{Parke:1996pr,Kodaira:1998gt}, $D_{LR}(\cos\theta)$ (and
$D_{RL}(\cos\theta)$) can be cast into a very compact Born term form 
\begin{equation}\label{dlr}
D_{LR}(Born)=\frac38\left(C_{LR}^2-2f_{LL}f_{LR}v^2\sin^2\theta\right)2N_cs,
\end{equation}
where
\begin{equation}\label{clr}
C_{LR}(\cos\theta)=f_{LL}(1+v\cos\theta)+f_{LR}(1-v\cos\theta)\,.
\end{equation}
The corresponding RL form $D_{RL}$ is obtained again by the substitution
$(L\leftrightarrow R)$ in Eqs.~(\ref{dlr}) and~(\ref{clr}). 

With the help of the compact expression in Eq.~(\ref{dlr}) and the translation
table $2(g_{11}-g_{41})=(f_{LL}^2+f_{LR}^2)$,
$2(g_{14}-g_{44})=-f_{LL}^2+f_{LR}^2$, $2(g_{11}+g_{41})=f_{RR}^2+f_{RL}^2$,
$2(g_{14}+g_{44})=f_{RR}^2-f_{RL}^2$ one can easily verify the threshold value
for $A_{LR}$ and the high energy limits for $A_{FB}$ discussed in
Sec.~\ref{sec2}. 

\section{\label{sec4}Single top polarization in $e^{+}e^{-} \to t \bar t$}
The polarization components $P^{(m)}$ ($m=\ell$: longitudinal; $m=tr$:
transverse) of the top quark in $e^{+}e^{-} \to t \bar t$ are obtained from
(the antitop quark spin is summed over)
\begin{equation}\label{pol}
P^{(m)}(P_{\rm eff})=\frac{N^{(m)}(P_{\rm eff})}{D(P_{\rm eff})}\,,
\end{equation}
where the dependence on $P_{\rm eff}$ is given by 
\begin{equation}
N^{(m)}(P_{\rm eff})=\frac14(1-h_-h_+)\left(N^{(m)}_{LR}+N^{(m)}_{RL}
  -P_{\rm eff}(N^{(m)}_{LR}-N^{(m)}_{RL})\right)\,.
\end{equation}
$P^{(tr)}$ is the transverse polarization component perpendicular to the
momentum of the top quark in the scattering plane. The overall helicity
alignment factor $(1-h_-h_+)$ drops out when one calculates the normalized 
polarization components according to Eq.~(\ref{pol}). This explains why the 
polarization depends only on $P_{\rm eff}$ and not separately on $h_-$ and 
$h_+$ (see Eq.~(\ref{vecpol})).

The numerator factors $N^{(m)}_{LR}$ and $N^{(m)}_{RL}$ in Eq.~(\ref{pol})
are given by
\begin{eqnarray}\label{lpol}
-2N^{(\ell)}_{LR}(\cos\theta)&=&\frac38(1+\cos^2\theta)\,
  (f_{LL}^2-f_{LR}^2)H_U^{4(\ell)}
  +\frac34\sin^2\theta\,(f_{LL}^2-f_{LR}^2)H_L^{4(\ell)}\nonumber\\&&
  +\frac34\cos\theta\,\bigg((f_{LL}^2+f_{LR}^2)H_F^{1(\ell)}
  +2f_{LL}f_{LR}H_F^{2(\ell)}\bigg)\,,
\end{eqnarray}
\begin{eqnarray}\label{trpol}
-2N^{(tr)}(\cos\theta)\!\!\!&=&\!\!\!
  -\frac3{\sqrt2}\sin\theta\cos\theta\,\,(f_{LL}^2-f_{LR}^2)\,H_I^{4(tr)}
  \nonumber\\&&
  -\frac3{\sqrt2}\sin\theta\left((f_{LL}^2+f_{LR}^2)H_A^{1(tr)}
  +2f_{LL}f_{LR}H_A^{2(tr)}\right)\,,
\end{eqnarray}
and $N^{(m)}_{RL}=-N^{(m)}_{LR}(L\leftrightarrow R)$. Note the extra minus 
sign when relating $N^{(m)}_{LR}$ and $N^{(m)}_{RL}$. 

The LO longitudinal and transverse polarization components read (see 
e.g.\ Ref.~\cite{Groote:2008ux,Groote:2010zf})
\begin{eqnarray}\label{ell}
H_U^{4(\ell)}({\it Born\/})=4N_csv,&&
H_F^{1(\ell)}({\it Born\/})=2N_cs(1+v^2),\nonumber\\[3pt]
H_L^{4(\ell)}({\it Born\/})=0,&&
H_F^{2(\ell)}({\it Born\/})=2N_cs(1-v^2),
\end{eqnarray}
and
\begin{equation}
H_I^{4(tr)}({\it Born\/})=2N_cs\frac{1}{2\sqrt{2}}v\sqrt{1-v^{2}},\quad
H_A^{1(tr)}({\it Born\/})=H_A^{2(tr)}({\it Born\/})
=2N_cs\frac{1}{2\sqrt{2}}\sqrt{1-v^2}\,.
\end{equation}

The LO  numerators~(\ref{lpol}) and~(\ref{trpol}) can be seen to take a
factorized form~\cite{Parke:1996pr,Kodaira:1998gt}
\begin{eqnarray}
N_{LR}^{(\ell)}(\cos\theta)&=&-\frac38\bigg(f_{LL}(\cos\theta+v)
  +f_{LR}(\cos\theta-v)\bigg)\,C_{LR}(\cos\theta)\,\,2N_cs\,,\label{psell}
\nonumber \\
N_{LR}^{(tr)}(\cos\theta)&=&\frac38\sin\theta\sqrt{1-v^2}\,
(f_{LL}+f_{LR})\,C_{LR}(\cos\theta)\,\,2N_cs\,,\label{pstr}
\end{eqnarray}
where the common factor $C_{LR}(\cos\theta)$ has been defined in
Eq.~(\ref{clr}). 

One can then determine the angle $\alpha$ enclosing the direction of the top
quark and its polarization vector by taking the ratio 
$N^{(tr)}_{LR}/N^{(\ell)}_{LR}$. One has
\begin{equation}\label{offdiag}
\tan\alpha_{LR}=\frac{N_{LR}^{(tr)}(\cos\theta)}{N_{LR}^{(\ell)}(\cos\theta)}
  =-\frac{\sin\theta\sqrt{1-v^2}\,(f_{LL}+f_{LR})}{f_{LL}(\cos\theta+v)
  +f_{LR}(\cos\theta-v)}\,.
\end{equation}
For $v=1$ one finds $\alpha_{LR}=0$, i.e.\ the polarization vector is aligned
with the momentum of the top quark, in agreement with what has been said 
before. In Ref.~\cite{Groote:2010zf} we have shown that radiative corrections
to the value of $\alpha_{LR}$ are small in the forward region but can become 
as large as $\Delta\alpha_{LR}=10^{\circ}$ in the backward region for large 
energies.

Eqs.~(\ref{psell}) and~(\ref{pstr}) can be used to find a very compact LO form
for $|\vec{P}_{LR}|$. One obtains~\cite{Groote:2010zf}
\begin{equation}\label{expanda}
|\vec{P}_{LR}|=\frac{\sqrt{N_{LR}^{(\ell)2}+N_{LR}^{(tr)2}}}{D_{LR}}
  =\frac{\sqrt{1-4a_{LR}}}{1-2a_{LR}}=1-2a_{LR}^2-8a_{LR}^3-18a_{LR}^3\ldots,
\end{equation}
where the coefficient $a_{LR}$ depends on $\cos\theta$ through
\begin{equation}
a_{LR}(\cos\theta)=\frac{f_{LL}f_{LR}}{C_{LR}^2(\cos\theta)}v^2\sin^2\theta\,.
\end{equation} 
Again, the corresponding expressions for $|\vec{P}_{LR}|$ and $a_{LR}$ can
be found by the substitution $(L\leftrightarrow R)$.

For the fun of it we also list a compact LO form for
$|\vec{P}(P_{\rm eff}=0)|$. One has
\begin{equation}\label{polpeff0}
|\vec{P}(P_{\rm eff}=0)|=\frac{\sqrt{(C_{LR}^2-C_{RL}^2)^2
  -4v^2\sin^2\theta(C_{LR}f_{LL}-C_{RL}f_{RR})(C_{LR}f_{LR}-C_{RL}f_{RL})}}
  {C_{LR}^2+C_{RL}^2-2v^2\sin^2\theta(f_{LL}f_{LR}+f_{RR}f_{RL})}\,.
\end{equation}
Eq.~(\ref{polpeff0}) would produce a LO version of Fig~\ref{fig:zeropol}. 

\section{\label{sec5}Effective beam polarization}
\begin{figure}[ht]
\begin{center}
\includegraphics[width=0.4\textwidth]{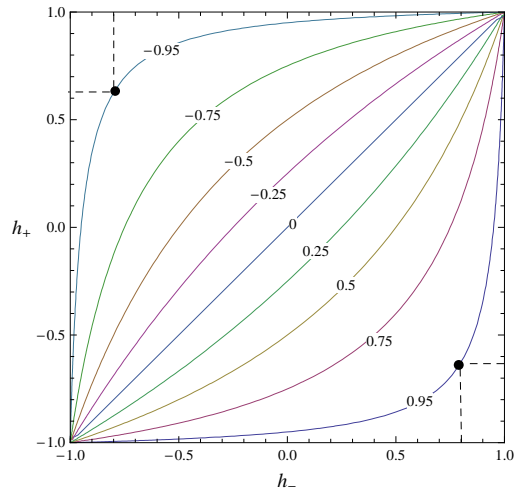} 
\end{center}
\vspace*{-0.5cm}
\caption{\label{fig:contour}Contour plots of $P_{\rm eff}={\it const}$ 
in the $(h_+,h_-)$ plane}
\end{figure}

As described in Sec.~\ref{sec2}, large values of the effective beam
polarization $P_{\rm eff}$ are needed to produce large polarization values of
$\vec{P}$. It is a fortunate circumstance that nearly maximal values of
$P_{\rm eff}$ can be achieved with non-maximal values of $(h_{-},h_{+})$. This
is shown in Fig.~\ref{fig:contour} where we drawn contour plots
$P_{\rm eff}={\it const}$ in the $(h_-,h_+)$ plane. The two examples shown in
Fig.~\ref{fig:contour} refer to
\begin{eqnarray}
&&(h_{-}=-0.80,\,h_{+}=+0.625)\qquad\mbox{leads to}\quad P_{\rm eff}=-0.95,
\nonumber\\
&&(h_{-}=+0.80,\,h_{+}=-0.625)\qquad\mbox{leads to}\quad P_{\rm eff}=+0.95.
\end{eqnarray} 
These two options are at the technical limits that can be
achieved~\cite{Alexander:2009nb}. In the next section we shall see that the
choice $P_{\rm eff}\sim-0.95$ is to be preferred since the polarization is
more stable against small variations of $P_{\rm eff}$. Furthermore, negative
values of $P_{\rm eff}$ gives yet another rate enhancement as discussed after
Eq.~(\ref{ratealr}).  

\section{\label{sec6}Stability of polarization against
  variations of $P_{\rm eff}$}
Extrapolations of $|\vec{P}|$ away from $P_{\rm eff}=\pm 1$ are more stable for
$P_{\rm eff}=-1$ than for $P_{\rm eff}=+1$. Because the derivative of the
magnitude of $|\vec{P}|$ leads to rather unwieldy expressions, we demonstrate
this separately for the two polarization components $P^{(\ell)}$ and
$P^{(tr)}$. The polarization components are given by ($m=\ell,tr$)
\begin{equation}
P^{(m)}=\frac{N_0^{(m)}-P_{\rm eff}N_P^{(m)}}{D_0-P_{\rm eff}D_P}\,,
\end{equation}
where $N_0^{(m)}=N_{LR}^{(m)}+N_{RL}^{(m)}$ and
$N_P^{(m)}=N_{LR}^{(m)}-N_{RL}^{(m)}$ and similarly for $D_0$ and $D_P$. Upon
differentiation w.r.t.\ $P_{\rm eff}$ one obtains
\begin{equation}
\frac{dP^{(m)}\,}{dP_{\rm eff}}=
\frac{-N_{0}^{(m)}D_{P}+N_{P}^{(m)}D_{0}}
{\left(D_{0}-P_{\rm eff}D_{P}\right)^{2}}\,.
\end{equation}

For the ratios of the slopes for $P_{\rm eff}=-1$ and $P_{\rm eff}=+1$ one
finds 
\begin{equation}
\frac{dP^{(m)}\,}{dP_{\rm eff}}\Big|_{P_{\rm eff}=-1}\,\bigg/\,
\frac{dP^{(m)}\,}{dP_{\rm eff}}\Big|_{P_{\rm eff}=+1}
  =\left(\frac{D_0-D_P}{D_0+D_P}\right)^2
  =\left(\frac{D_{RL}}{D_{LR}}\right)^2
  =\left(\frac{1-A_{LR}}{1+A_{LR}}\right)^2\,.
\end{equation}
Depending on the energy and the scattering angle, Fig.~\ref{fig:asymrl} shows
that $A_{LR}$ varies between $0.3$ and $0.7$ which implies that
$(D_{RL}/D_{LR})^2$ varies between $0.29$ and $0.06$, i.e.\ for
$P_{\rm eff}=-1$ the polarization components are much more stable against
variations of $P_{\rm eff}$ than for $P_{\rm eff}=+1$. At threshold the ratio
of slopes of $|\vec{P}_{\rm thresh}\,|$ for $P_{\rm eff}=-1$ and
$P_{\rm eff}=+1$ is given by $-(D_{RL}/D_{LR})^2=-0.18$ where the minus sign
results from having taken the derivative of the magnitude $|\vec{P}\,|$ (see
Eq.~(\ref{thresh})).

\section{\label{sec7}Longitudinal and transverse polarization\\
  $P^{(\ell)}$ vs.\ $P^{(tr)}$ for general angles and energies}
\begin{figure}[ht]
\begin{center}
\begin{tabular}{lr}
\includegraphics[width=0.4\textwidth]{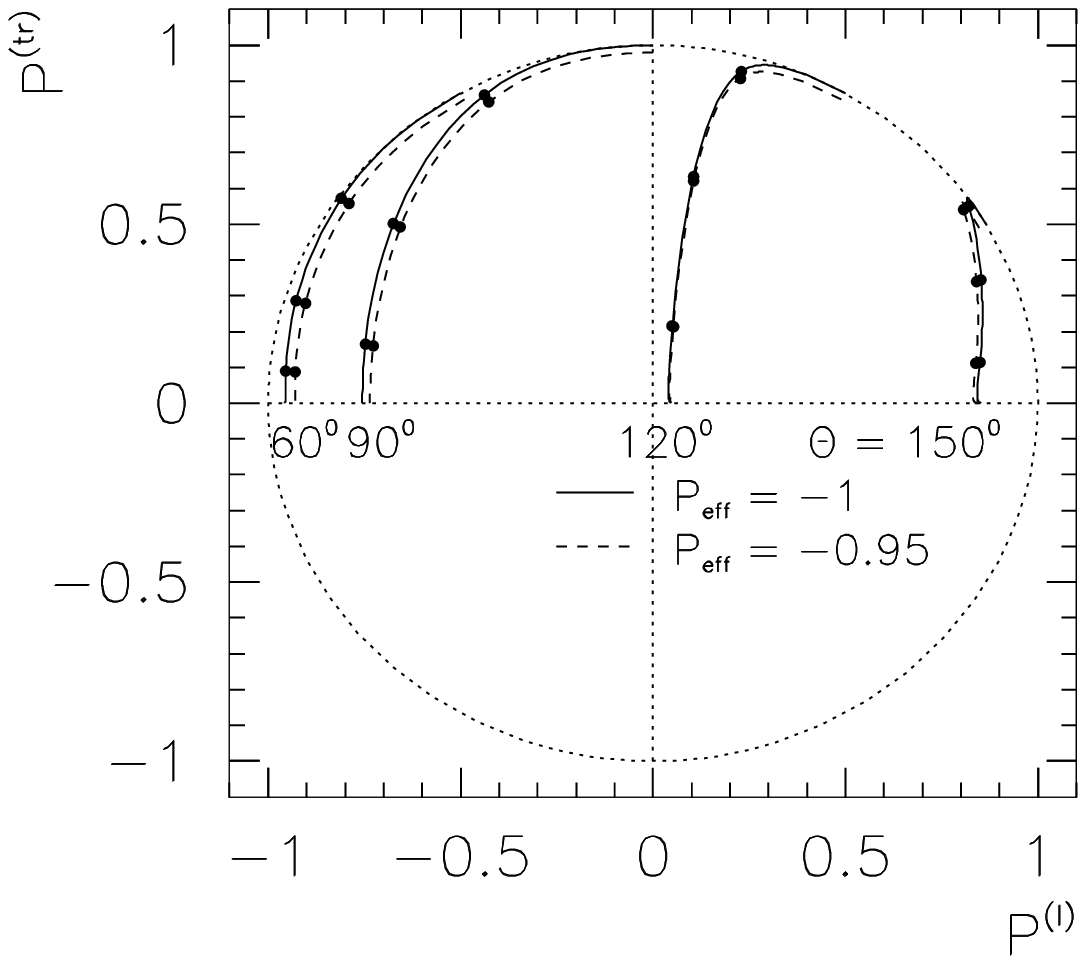} &
\includegraphics[width=0.4\textwidth]{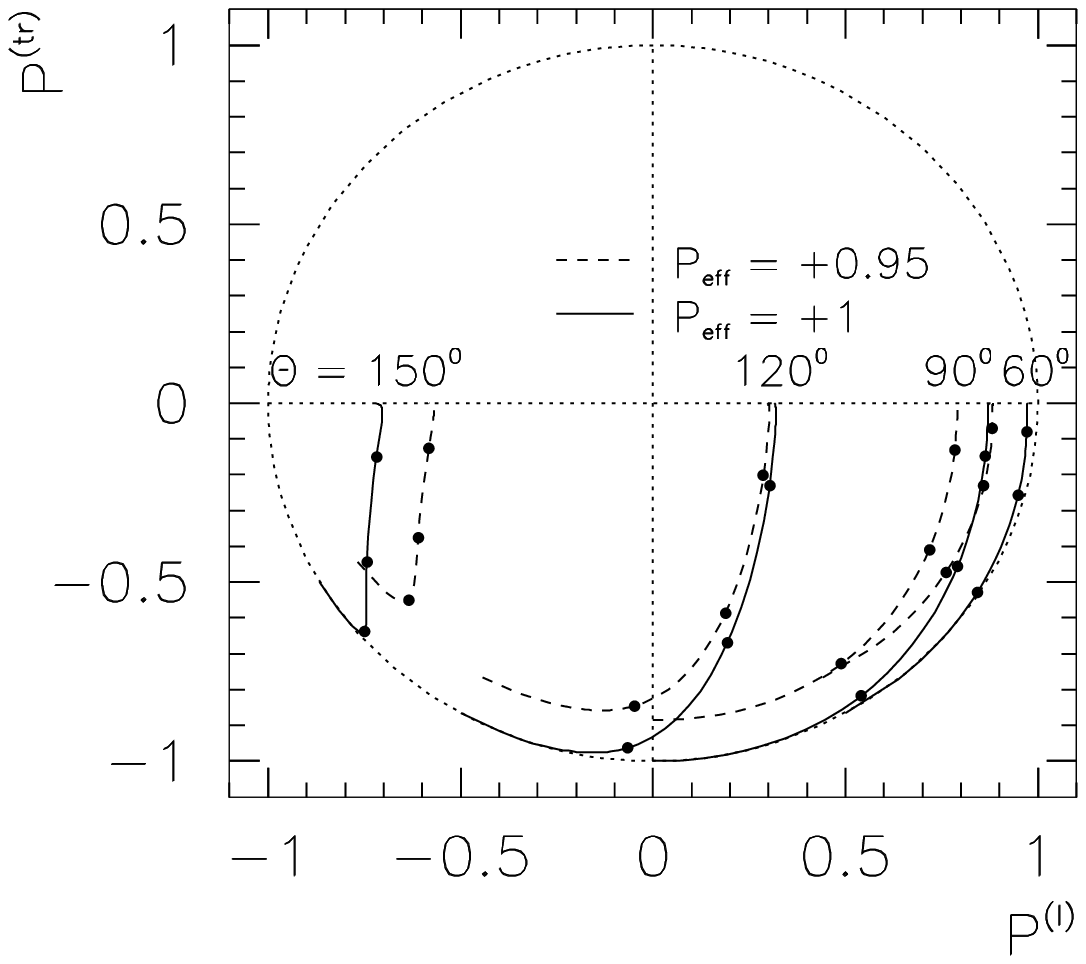} \\[2ex]
\end{tabular}
\end{center}
\vspace*{-0.5cm}
\caption{\label{fig:translong1}Parametric plot of the orientation and the
  length of the polarization vector in dependence on the c.m.\ energy
  $\sqrt{s}$ for values $\theta=60^\circ$, $90^\circ$, $120^\circ$, and
  $150^\circ$ for i) (left panel) $P_{\rm eff}=-1$ (solid lines) and
  $P_{\rm eff}=-0.95$ (dashed lines) and ii) (right panel) $P_{\rm eff}=+1$
  (solid lines) and $P_{\rm eff}=+0.95$ (dashed lines). The three ticks on the
  trajectories stand for $\sqrt{s}=500\,GeV$, $1000\,GeV$, and $3000\,GeV$.}
\end{figure}

In Fig.~\ref{fig:translong1} we plot the longitudinal component $P^{(\ell)}$
and the transverse component $P^{(tr)}$ of the top quark polarization for
different scattering angles $\theta$ and energies $\sqrt{s}$ starting from
threshold up to the high energy limit. The left and right panels of
Fig.~\ref{fig:translong1} are drawn for $P_{\rm eff}=(-1,-0.95)$ and for
$P_{\rm eff}=(+1,+0.95)$, respectively. The apex of the polarization vector
$\vec{P}$ follows a trajectory that starts at
$\vec{P}=P_{\rm thresh}(-\cos\theta,\sin\theta)$ and
$\vec{P}=P_{\rm thresh}(\cos\theta,-\sin\theta)$ for negative and positive
values of $P_{\rm eff}$, respectively, and ends on the line $P^{(tr)}=0$ in
the high energy limit. The two $60^{\circ}$ trajectories show that large
values of the size of  $|\vec{P}|$ close to the maximal value of $1$ can be
achieved in the forward region for both $P_{\rm eff}\sim\mp 1$ at all energies.
However, the two figures also show that the option $P_{\rm eff}\sim -1$ has to
be  preferred since the $P_{\rm eff}\sim-1$ polarization is more stable
against variations of $P_{\rm eff}$.

It is noteworthy that the magnitude of the polarization vector remains closer
to $|\vec{P}|=1$ in the forward region than in the backward region when
$\cos\theta$ is varied. Let us investigate this effect for $P_{\rm eff}=-1$ by
expanding the high energy formula~(\ref{helimit}) in $\Delta\cos\theta$ around
$\cos\theta=+1$ and $\cos\theta=-1$. Since the first derivative vanishes, one
has to expand to the second order in $\Delta\cos\theta$. The result is
\begin{eqnarray}
{\rm Forward}\qquad|\vec{P}_{LR}|
  &=&1-\frac12\left(\frac{f_{LR}}{f_{LL}}\right)^2(\Delta\cos\theta)^2
  +\ldots\nonumber\\
{\rm Backward}\qquad|\vec{P}_{LR}|
  &=&1-\frac12\left(\frac{f_{LL}}{f_{RL}}\right)^2(\Delta\cos\theta)^2
  +\ldots\,.
\end{eqnarray} 

Numerically, one has $f_{LR}^2/f_{LL}^2=0.13$ and $f_{LL}^2/f_{LR}^2=7.53$.
The second derivative is very much smaller in the forward direction than in
the backward direction. This tendency can be clearly discerned in
Fig.~\ref{fig:translong1}. A similar but even stronger conclusion is reached
for the second derivative of $|\vec{P}_{RL}|$ where the corresponding second
order coefficients are given by $f_{RL}^2/f_{RR}^2=0.064$ for $\cos\theta=+1$,
and by $f_{RR}^2/f_{RL}^2=15.67$ for $\cos\theta=-1$. Corresponding
$v$-dependent expansions can be obtained from Eq.~(\ref{expanda}).

We mention that at NLO there is also a normal component of the top quark 
polarization $P^{(n)}$ generated by the one--loop contribution which, however,
is quite small (of $O(3\%))$ \cite{Groote:2010zf}.

\section{\label{sec8}Summary}
The aim of our investigation was to maximize and to minimize the polarization
vector of the top quark $\vec{P}\,(\sqrt{s},\cos\theta,g_{ij},P_{\rm eff})$
by tuning the beam polarization. Let us summarize our findings which have been
found in NLO QCD in the context of the SM.\\[12pt]
{\bf A. Maximal polarization:}
Large values of $\vec{P}$ can be realized for $P_{\rm eff}\sim\pm1$ at all
intermediate energies. This is particularly true in the forward hemisphere
where the rate is highest. Negative large values for $P_{\rm eff}$ with
aligned beam helicities ($h_-h_+$ neg.) are preferred for two reasons. First
there is a further gain in rate apart from the helicity alignment factor
$(1-h_-h_+)$ due to the fact that generally $\sigma_{LR}>\sigma_{RL}$ as
explained after Eq.~(\ref{rate}). Second, the polarization is more stable
against variations of $P_{\rm eff}$ away from $P_{\rm eff}=-1$. The forward
region is also favoured since the $100\%$ LO polarization valid at 
$\cos\theta=1$ extrapolates smoothly into the forward hemisphere with small
radiative corrections. \\[12pt]
{\bf B. Minimal polarization:}
Close to zero values of the polarization vector $\vec{P}$ can be achieved for
$P_{\rm eff}\sim 0.4$. Again the forward region is favoured. In order to
maximize the rate for the small polarization choice take quadrant IV in the
$(h_-,\,h_+)$ plane.


\section*{Acknowledgements}
J.G.K.\ would like to thank X.~Artru and E.~Christova for discussions and
G.~Moortgat-Pick for encouragement. The work of S.G.\ is supported by the
Estonian target financed project No.~0180056s09, by the Estonian Science
Foundation under grant No.~8769 and by the Deutsche Forschungsgemeinschaft
(DFG) under grant 436 EST 17/1/06. B.M.\ acknowledges support of the Ministry
of Science and Technology of the Republic of Croatia under contract
No.~098-0982930-2864. S.P.\ is supported by the Slovenian Research Agency. 

\begin{footnotesize}

\end{footnotesize}

\end{document}